\documentclass[12pt]{article}
\usepackage{amssymb}
\def\Rset{\mathbb{R}}

\def\Rset{\mathbb{R}}
\newcommand{\refbr}[1]{(\ref{#1})}
\newcommand{\beq}{\begin{equation}}
\newcommand{\eeq}{\end{equation}}

\newcommand{\norm}[1]{|| #1 ||}

\def\sf{\sffamily}

\begin{document}

\title{
{\sf {\small Submitted to "IEEE/ACM Transactions on Networking" in August 2002}}\\
\vskip 2cm
Correction to Low and Lapsley's
article "Optimization Flow Control, I: Basic Algorithm and Convergence" }
\author{Andrzej Karbowski\\
NASK (Research and Academic Computer Network),\\
ul. W¹wozowa 18,\\
02-796~Warsaw, Poland,\\
E-mail: A.Karbowski@ia.pw.edu.pl}

\date{ }
\parindent 2em
\maketitle

\vskip 2cm
\begin{abstract}

In the note an error in Low and Lapsley's article \cite{Low1} is
pointed out.
Because of this error the proof of the Theorem 2 presented in the article
is incomplete and some assessments are wrong.
In the second part of the note the author proposes
a correction to this proof.

\end{abstract}

\newpage
\section{Error in the proof of the Theorem 2}

The last passage in the assessment (29) on the page 873 (the proof of Lemma 6),
namely
\[
D(p(t+1)) \leq   D(p(t)) - \left(\frac{1}{\gamma} - A_1\right) \norm{\pi(t)}^2
\]
\[
+ A_2 \sum_{t'=t-2t_0}^{t-1} \norm{ \pi(t')} \cdot \norm{\pi(t)}\\
\]
\[
\leq D(p(t)) - \left(\frac{1}{\gamma} - A_1\right)\norm{\pi(t)}^2
\]
\beq
+ A_2  \sum_{t'=t-2t_0}^{t} \norm{\pi(t)}^2
\label{ineq.29}
\eeq
is incorrect.
Low and Lapsley justify it: {\em "where the last inequality holds because the convex
function $\sum_i y_i^2 + z^2 - \sum_i  y_i  z$ attains a unique
minimum over $\{(y_i,z) \,| \;y_i \geq 0, z  \geq 0 \}$ at the origin."}\\
But this is not true.
It is sufficient to take $\dim y = 5$ and $ y = [5,\, 4,\, 3,\, 4,\, 5]$, $ z = 10$.
The considered function takes the value $-19$.\\
It is so, because the Hessian of this function has the form
(taking: $x~=~[y_1, y_2, \ldots, z]'$):
\beq
H = \left[ \begin{array}{cccccc}
 2 & 0 &0 &\ldots &0 & -1 \\
 0 & 2 & 0 &\ldots & 0 &-1 \\
 0 & 0 & 2 &\ldots & 0 &-1 \\
 \vdots\\
  0 & 0 & 0 &\ldots & 2 &-1 \\
  -1 & -1 & -1 & \ldots& -1 & 2
 \end{array}\right]
 \eeq
and its characteristic polynomial  (e.g. calculated from the
Schur's formula: for $A~:~n~\times~n;$\, $D~:~m~\times~m,\; B^T,C:m \times n; \;det([A, B; C, D]) =  det(A) \cdot det(D-CA^{-1}B)$\, )  will be:
\beq
det(H - \lambda I) = (2 - \lambda)^{n-1} (\lambda^2 - 4 \lambda + 4 -n)
\eeq
where $n = \dim y$. In this way we will have  eigenvalues: $\lambda_i = 2,\, i=1,\ldots,n-1$
and $\lambda_{n,n+1}~=~2~\pm~\sqrt{n}$. This means, that for $n>4$ there will be one
negative eigenvalue and the function will not  be convex.

\section{Correction to the proof of the Theorem 2}

The mentioned passage in the assessment (29) in \cite{Low1} (here ineq. \refbr{ineq.29})
should be changed. The changes are based on a very simple assessment:
\beq
\forall \,y_i, z \in \Rset \;\;\;\; (y_i-z)^2=y_i^2-2y_i z+z^2 \geq 0
\eeq
From which after elementary operations we get:
\beq
y_i z \leq \frac{y_i^2+z^2}{2}
\label{oszac_i}
\eeq
Let us apply the assessment \refbr{oszac_i} to all elements of a finite set
of real numbers $\{y_i, i \in I\}$, where $I$ is a set of integer indices,
and sum up both sides of these inequalities over all $i \in I$.
We will obtain:
\beq
\sum_{i \in I} y_iz \leq \frac{1}{2} \sum_{i \in I} \left(y_i^2+z^2\right)
=\frac{1}{2}\,\overline{\overline{I}}\cdot z^2+\frac{1}{2} \sum_{i \in I} y_i^2
\eeq
That is:
\beq
\sum_{i \in I} y_iz \leq
\frac{1}{2}\,\overline{\overline{I}}\cdot z^2+\frac{1}{2} \sum_{i \in I} y_i^2
\label{oszac_kon}
\eeq
where $\overline{\overline{I}}$ is the number of elements of the set $I$.
We will use the assessment \refbr{oszac_kon} to transform the first part of assessment
\refbr{ineq.29}.
In particular, owing to \refbr{oszac_kon}, for the last component of
the right hand side we will have:
\[
\sum_{t'=t-2t_0}^{t-1}\norm{\pi(t')} \cdot\norm{\pi(t)} \leq
\frac{1}{2} \left[t-1-\left(t-2t_0\right)+1\right] \cdot \norm{ \pi(t)}^2 +
\frac{1}{2}\sum_{t'=t-2t_0}^{t-1}\norm{\pi(t')}^2 =
\]
\beq
=t_0 \norm{ \pi(t)}^2 +\frac{1}{2}\sum_{t'=t-2t_0}^{t-1}
\norm{\pi(t')}^2
\eeq

Let us notice that:
\[
t_0 \norm{ \pi(t)}^2 +\frac{1}{2}\sum_{t'=t-2t_0}^{t-1}
\norm{\pi(t')}^2
=t_0 \norm{ \pi(t)}^2 -
\frac{1}{2}  \norm{ \pi(t)}^2 +\frac{1}{2} \norm{ \pi(t)}^2+
\]
\beq
 +\frac{1}{2} \sum_{t'=t-2t_0}^{t-1}
\norm{\pi(t')}^2 = \left(t_0-\frac{1}{2}\right)\norm{ \pi(t)}^2
+\frac{1}{2} \sum_{t'=t-2t_0}^{t} \norm{\pi(t')}^2
\label{suma_29}
\eeq
So, for the value of $\sum_{t'=t-2t_0}^{t-1}\norm{\pi(t')} \cdot\norm{\pi(t)}$ we will
have the following assessment:
\beq
\sum_{t'=t-2t_0}^{t-1}\norm{\pi(t')} \cdot\norm{\pi(t)}
\leq \left(t_0-\frac{1}{2}\right)\norm{ \pi(t)}^2
+\frac{1}{2} \sum_{t'=t-2t_0}^{t} \norm{\pi(t')}^2
\eeq
The correct form of the assessment (29) in the article \cite{Low1} will be then:
\[
D(p(t+1)) \leq   D(p(t)) - \left(\frac{1}{\gamma} - A_1\right) \norm{\pi(t)}^2
\]
\[
+ A_2 \sum_{t'=t-2t_0}^{t-1} \norm{ \pi(t')}\norm{\pi(t)}\\
\]
\[
\leq D(p(t)) - \left(\frac{1}{\gamma} - A_1\right)\norm{\pi(t)}^2
\]
\[
+ A_2 \left(t_0-\frac{1}{2}\right)\norm{ \pi(t)}^2
+\frac{A_2}{2} \sum_{t'=t-2t_0}^{t} \norm{\pi(t')}^2 =
\]
\[
= D(p(t)) - \left[\frac{1}{\gamma} - A_1 - A_2 \left(t_0-\frac{1}{2}\right)\right]\norm{\pi(t)}^2
+\frac{A_2}{2} \sum_{t'=t-2t_0}^{t} \norm{\pi(t')}^2
\]
That is:
\beq
D(p(t+1)) \leq D(p(t)) - \left[\frac{1}{\gamma} - A_1 - A_2 \left(t_0-\frac{1}{2}\right)\right]\norm{\pi(t)}^2
+\frac{A_2}{2} \sum_{t'=t-2t_0}^{t} \norm{\pi(t')}^2
\label{fin.assess}
\eeq
Now, applying this inequality recursively to all $D(p(\tau)),\; \tau=t,t-1,\ldots,1$,
taking $\pi(t)=0$ for $t <0$, we will get the following total assessment which replaces wrong assessment (30) in the
article \cite{Low1}:
\beq
D(p(t+1)) \leq D(p(0)) - \left[\frac{1}{\gamma} - A_1 - A_2 \left(t_0-\frac{1}{2}\right)\right]
\sum_{\tau=0}^{t}\norm{\pi(\tau)}^2
+\frac{A_2}{2}\sum_{\tau=0}^{t} \sum_{t'=\tau-2t_0}^{\tau} \norm{\pi(t')}^2
\label{zam.30}
\eeq
Let us notice, that under the assumption that $\pi(t)=0$ for $t <0$:
\[
\sum_{\tau=0}^{t} \sum_{t'=\tau-2t_0}^{\tau} \norm{\pi(t')}^2
= \sum_{\tau=0}^{t} \sum_{t''=0}^{2t_0} \norm{\pi(t''+\tau-2t_0)}^2
= \sum_{t''=0}^{2t_0} \sum_{\tau=0}^{t} \norm{\pi(\tau+t''-2t_0)}^2
\]
\beq
\leq \sum_{t''=0}^{2t_0} \sum_{\tau=0}^{t} \norm{\pi(\tau)}^2 =
(2t_0+1) \sum_{\tau=0}^{t} \norm{\pi(\tau)}^2
\label{dwie.sumy}
\eeq
Putting \refbr{dwie.sumy} into \refbr{zam.30} we finally get:
\[
D(p(t+1)) \leq D(p(0)) - \left[\frac{1}{\gamma} - A_1 - A_2 \left(t_0-\frac{1}{2}\right)\right]
\sum_{\tau=0}^{t}\norm{\pi(\tau)}^2
+\frac{A_2}{2}(2t_0+1) \sum_{\tau=0}^{t} \norm{\pi(\tau)}^2
\]
\[
=D(p(0)) - \left[\frac{1}{\gamma} - A_1 - A_2 \left(t_0-\frac{1}{2}\right)
-A_2\left(t_0+\frac{1}{2}\right) \right]
\sum_{\tau=0}^{t}\norm{\pi(\tau)}^2 =
\]
\beq
D(p(0)) - \left(\frac{1}{\gamma} - A_1 - 2 A_2 t_0\right)
\sum_{\tau=0}^{t}\norm{\pi(\tau)}^2
\eeq
That is:
\beq
D(p(t+1))\leq D(p(0)) - \left(\frac{1}{\gamma} - A_1 - 2 A_2 t_0\right)
\sum_{\tau=0}^{t}\norm{\pi(\tau)}^2
\label{my.30}
\eeq

\end{document}